\documentclass{ws-p9-75x6-50}
\usepackage{graphicx}  
\begin{document}

\title{Dissipation in Open Two--Level Systems: Perturbation Theory and Polaron Transformation}

\author{T. Brandes}

\address{Dept. of Physics, UMIST, Manchester M60 1QD, UK\\E-mail: Tobias.Brandes@umist.ac.uk}

\author{T. Vorrath}

\address{1. Institut f\"ur Theoretische Physik, Universit\"at Hamburg, Jungiusstr. 9, 20355 Hamburg, Germany}


\maketitle

\abstracts{We compare standard perturbation theory with the polaron transformation for non--linear transport of 
electrons through a two--level system.  For weak
electron--phonon coupling and large bias, there is good agreement between both approaches.
This regime has recently been explored in experiments in double quantum dots.}

\section{Introduction}
Semiconductor coupled quantum dots 
are well--defined artificial few--particle systems for the study of
electron--electron {correlations}, {quantum coherence}, and {quantum dissipation}.
These effects become visible in non--linear electron transport on an $\mu$eV 
energy scale when parameters like the interdot coupling and the coupling to external leads
can be controlled by, e.g., external gate--voltages \cite{Fujetal98}.

Double quantum dots in the regime of strong Coulomb blockade
can be tuned into a regime that is governed by a (pseudo) spin--boson model \cite{BK99}
(dissipative two--level system)
\begin{eqnarray}\label{modelhamiltonian}
  H &=& \frac{\varepsilon}{2}\sigma_z + T_c \sigma_x+
\frac{1}{2}\sigma_z A + \sum_{\bf Q}\omega_{Q} a^{\dagger}_{\bf Q}
a_{\bf Q},\quad
A:= \sum_{\bf Q} g_{Q} \left(a_{-\bf Q} + a^{\dagger}_{\bf Q}\right),
\end{eqnarray}
where one additional `transport' electron tunnels between a left (L) and a right (R) 
dot with energy difference $\varepsilon$ and inter--dot coupling  $T_c$, where
$\sigma_z=|L \rangle \langle L|-|R \rangle \langle R|$ and
$\sigma_x=|L \rangle \langle R|+|R \rangle \langle L|$. Here,
$\omega_Q$ are the frequencies of phonons, and
the $g_Q$ denote interaction constants.
Although not exactly solvable, the model is quite well understood for closed systems \cite{Legetal87}
(isolated dots with one additional electron).
The coupling to external leads offers the possibility to study its
non--equilibrium properties, such as the inelastic stationary current through the dots.

\section{Equations of Motion}
We describe the dynamics of the double dot 
by a reduced statistical operator $\rho(t)$, allowing for an additional `empty' state and 
tunneling from a left reservoir at rate $\Gamma_L$ into the left dot, and from the right dot to
the right reservoir at rate  $\Gamma_R$. Lowest order perturbation theory in these rates
yields \cite{BK99,SN96}
\begin{eqnarray}
  \frac{\partial}{\partial t}\rho_{LL}(t)&=&-iT_c\left[
\rho_{LR}(t)-\rho_{RL}(t) \right] 
+
\Gamma_L\left[1-\rho_{LL}(t)-\rho_{RR}(t)\right]
\nonumber\\
\frac{\partial}{\partial t}\rho_{RR}(t)&=&-iT_c\left[
\rho_{RL}(t)-\rho_{LR}(t) \right] 
-{\Gamma}_R\rho_{RR}(t)\nonumber.
\end{eqnarray}
For the remaining equation for the off--diagonal element $\rho_{LR}=\rho_{RL}^*$, one has to choose
between perturbation theory in $g_Q$ (weak coupling, PER), or in $T_c$ in a polaron--transformed
frame (strong coupling, POL). In general, no exact solution of the model is available: this is the case even for
coupling to one bosonic mode only ($g_{Q} \propto \delta_{Q,Q_0}$, Rabi Hamiltonian). 

For the spin--boson problem with $\Gamma_{R/L}=0$, 
it is well--known that POL is equivalent to a double--path integral `non--interacting blip approximation' (NIBA) that works
well for zero bias $\varepsilon=0$ but for $\varepsilon\ne 0$  does not coincide with PER at small couplings 
and low temperatures. In the following, we compare both approaches for $\Gamma_{R/L}\ne 0$ and find 
nearly perfect agreement for very large $\varepsilon \gg T_c$, a regime that has been tested experimentally 
recently \cite{Fujetal98}.

The standard Born and Markov approximation with respect to  $A$ yields
\begin{eqnarray}
\frac{d}{dt}\rho_{LR}^{\rm PER}(t) 
&=& \left[i\varepsilon - \gamma_p - \Gamma_R/2 \right]
\rho_{LR}(t) 
+ \left[i T_c -\delta_-\right]
\rho_{RR}(t)-\left[iT_c-\delta_+ \right]\rho_{LL}(t).\nonumber
\end{eqnarray}
Here, the rates are 
\begin{eqnarray}
  \gamma_p&:=& 2\pi \frac{T_c^2}{\Delta^2} \rho (\Delta) \coth \left(\beta \Delta /2\right),
\quad
\delta_{\pm}:= -\frac{\varepsilon T_c}{\Delta^2}
\frac{\pi}{2} \rho (\Delta)  \coth \left(\beta \Delta /2\right)\mp 
 \frac{T_c}{\Delta} \frac{\pi}{2}\rho (\Delta)\nonumber\\
\rho({\omega})&:=&\sum_{\bf Q} |g_{Q}|^2\delta(\omega-\omega_Q).
\end{eqnarray}
where $\Delta:=\sqrt{\varepsilon^2+4T_c^2}$ is the energy
difference of the hybridized levels, and $\beta=1/k_BT$ the inverse phonon equilibrium bath temperature.
Note that beside the off--diagonal decoherence rate $\gamma_{p}$, there appear
terms $\propto \delta_{\pm}$ in the diagonals which below turn out to be important
for the stationary current.

On the other hand, the polaron transformation \cite{BK99} leads to an integral equation
\begin{eqnarray}\label{eom3new}
\rho_{LR}^{POL}(t)&=& -\int_0^tdt'   e^{i\varepsilon(t-t')}
\Bigg [ \frac{{\Gamma}_R}{2}C(t-t')
\rho_{LR}(t') \nonumber\\
&+ &
iT_c  \left\{ C(t-t') \rho_{LL}(t') - C^*(t-t')
\rho_{RR}(t')\right\} \Bigg ], \nonumber
\end{eqnarray}
where
\begin{eqnarray}
  C(t):= \exp\left\{ -\int_0^{\infty}d\omega \frac{\rho(\omega)}{\omega^2} 
\left[ \left(1- \cos \omega t\right) \coth (\beta \omega /2)
+ i \sin \omega t \right]\right\}.
\end{eqnarray}

\section{Stationary Current}
If an electron tunnels
between two lateral dots, the interaction with 3d piezoelectric acoustic phonons leads to an ortho\-go\-nality
catastrophe ({`boson shake up effect'}) 
which is
determined by an effective phonon density of states \cite{BK99}
\begin{equation}
  \rho(\omega)= 
{g}{\omega}\left(1-\frac{\omega_d}{\omega}\sin\left(\frac{\omega}{\omega_d}\right)\right)
 e^{-\omega/\omega_c}\theta(\omega),
\end{equation}
showing oscillations on a scale $\omega_d:=c_s/d$, where $c_s$ is the speed of sound, $d$ the
distance between the centers of the two dots, $g$ the dimensionless coupling constant, and
$\omega_c$ a high--energy cut--off. We argue that  the assumption of sharply localized 
positions between which the additional electron tunnels is justified by the 
strong intra--dot electron--electron repulsion.

We compare results for the stationary electron current $I_{\rm stat}=-e2T_c {\rm Im}
\hat{\rho}_{LR}(z=0)$ as obtained  from Laplace transforming the equations of motion 
within both PER and POL for small electron--phonon coupling. The result
\begin{eqnarray}\label{current}
  I_{\rm stat} &=& 
\frac{-e\Gamma_L\Gamma_RG_+}
{\Gamma_LG_- + (\Gamma_L+\Gamma_R)G_+
- \Gamma_L\Gamma_{R}}.
\end{eqnarray}
has to be used with either 
\begin{eqnarray}
G_{\pm}^{(\rm PER)}&:=&2T_c {\rm Im} \frac{iT_c -\delta_{\pm} }{i\varepsilon
- \gamma_p - \Gamma_R/2}
\end{eqnarray}
or 
\begin{eqnarray}
G_{+}^{(\rm POL)}&:=&
2T_c {\rm Im} \frac{-iT_cC_{\varepsilon}}{1+
(1/2){\Gamma}_RC_{\varepsilon}},\quad
G_{-}^{(\rm POL)}:=
2T_c {\rm Im} \frac{-iT_cC_{-\varepsilon}^*}{1+
(1/2){\Gamma}_RC_{\varepsilon}},
\end{eqnarray}
where $C_{\varepsilon}:=\int_0^{\infty}dte^{i\varepsilon t} C(t)$. 

The function $C_{\varepsilon}$ is obtained for arbitrary large coupling constant $g$ by a double
numerical integration. For our purpose here, it is sufficient to expand $C_{\varepsilon}$ as
\begin{eqnarray}
C_{\varepsilon}=\frac{i}{\varepsilon} + \gamma(\varepsilon)+O(g^2),\quad 
\gamma(\varepsilon):=
\frac{\pi}{2} \frac{\rho(|\varepsilon|)}{\varepsilon^2}\left[\coth(\beta|\varepsilon|/2)
+{\rm sgn }(\varepsilon)\right].
\end{eqnarray}

\section{Discussion}
In both PER and POL the current Eq. (\ref{current}) is an infinite sum of 
contributions $G_+$ ($=I_{\rm stat}/e$ in lowest order in $T_c$) and $G_-$ and therefore
to infinite order in the inter--dot coupling $T_c$. 
Note that both expressions for $G_{\pm}$
coincide for vanishing electron--phonon coupling $g$.
 
We first point out that PER works in the correct eigenstate base of the hybridized system
(level splitting $\Delta$), whereas the energy
scale $\varepsilon$ in POL is that of the two isolated dots ($T_c=0$). 
This difference reflects the general dilemma of two--level--boson Hamiltonians: either one is 
in the correct base of the hybridized two--level system and perturbative in $g$, or
one starts from the `shifted oscillator' polaron picture that becomes correct for
$T_c=0$. In fact, the polaron (NIBA) approach does not coincide with
standard damping theory \cite{SW90} because it does not
incorporate the  square--root, non--perturbative in $T_c$
hybridization form of $\Delta= \sqrt{\varepsilon^2+4T_c^2}$.

However, for large $|\varepsilon| \gg T_c$,
$\Delta \to |\varepsilon|$, and POL and PER should coincide. 
This is indeed the case for small $g$ where
\begin{eqnarray}
 -G_{\pm}\to 2T_c^2\frac{\Gamma_R/2+\gamma(\pm \varepsilon)}{\Gamma^2_{R}/4+\varepsilon^2}. 
\end{eqnarray}
Thus, for large  $|\varepsilon|$, the polaron approach
works well even down to very low temperatures and small coupling constants.
In particular, in the spontaneous emission regime 
of large positive $\varepsilon$ 
the agreement is very good. This regime is most interesting at low temperatures
$T$ and has been tested in the non--linear transport experiment \cite{Fujetal98}
in great detail.
\begin{figure}[th]
\includegraphics[width=0.7\textwidth]{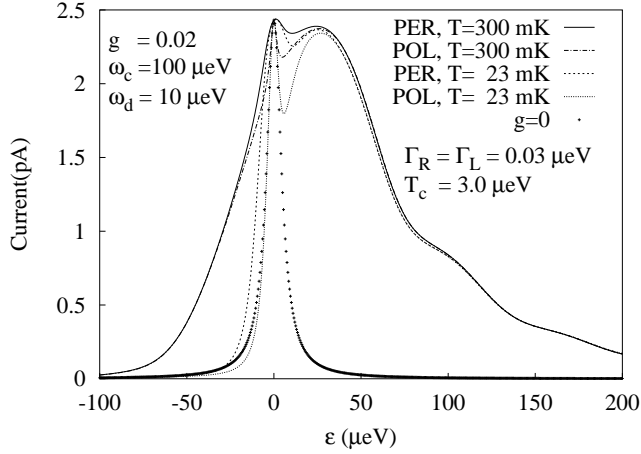}
\caption{\label{current_nl.eps}
Stationary current through double dot with energy difference
$\varepsilon$ for constant tunnel coupling $T_c$.}
\end{figure}

In Fig. 1, we compare both approaches and show the calculated stationary current,
using realistic parameters. 
In the vicinity of $\varepsilon =0$, the agreement between POL and PER
gets worse as could be expected 
from the above argument. At higher temperatures, the agreement  
gets slightly better  on the absorption side $\varepsilon<0$.
We conclude that our findings for the `open' spin--boson model are in agreement 
with standard spin--boson physics. Furthermore,
we see that for  large bias $|\varepsilon| \gg T_c$ both perturbation theory 
and the polaron transformation approach practically coincide for small coupling $g$.

This work was supported by the DFG, Project BR 1528/4, and the EPSRC, GR/R44690/01.

\end{document}